\documentclass[a4paper,fleqn,usenatbib]{mnras}
\usepackage{mathptmx}
\usepackage[T1]{fontenc}
\usepackage{ae,aecompl}
\usepackage{amssymb,amsmath,amsfonts,graphicx}
\usepackage{natbib,color}

\newcommand{\lcdm}{{$\Lambda$CDM}}
\newcommand{\Msun}{{$\rm M_\odot$}}
\newcommand{\de}{{\rm d}}

\begin{document}

\title[GCs in dark halos]{Globular Clusters Formed within Dark Halos I: present-day abundance,
distribution and kinematics}
\author[P. Creasey et al.]{Peter Creasey$^{1}$\thanks{E-mail: \href{mailto:peter.creasey@ucr.edu}{peter.creasey@ucr.edu}}, 
Laura V. Sales$^1$, Eric W. Peng$^2$, and Omid Sameie$^1$\thanks{NASA MIRO FIELDS fellow} \\
$^1$ Department of Physics and Astronomy, University of California, Riverside, California 92507, USA \\
$^2$ KIAA \& Department of Astronomy, Peking University, Beijing 100871, China}

\maketitle

\begin{abstract} 
  We explore a scenario where metal poor globular clusters (GCs) form
  at the centres of their own dark matter halos in the early Universe
  before reionization. This hypothesis leads to 
  predictions about the abundance, distribution and kinematics of GCs
  today that we explore using cosmological N-body simulations and
  analytical modelling. We find that selecting the massive tail of
  collapsed objects at $z\gtrsim 9$ as GCs formation sites leads to four main
  predictions: $i)$ a highly clustered population of GCs around
  galaxies today, $ii)$ a natural scaling between number of GCs and
  halo virial mass that follows roughly the observed trend $iii)$ a
  very low number of free floating GCs outside massive halos and $iv)$
  GCs should be embedded within massive and extended dark
  matter (sub)halos. We find
  that the strongest constraint to the model is given by the
  combination of $i)$ and $ii)$: a mass cut to tagged GCs halos which
  accounts for the number density of metal poor GCs today predicts a
  radial distribution that is too extended compared to recent
  observations. On the other hand, a mass cut sufficient
  to match the observed half number radius could only explain $60\%$
  of the metal poor population. In all cases, observations favour early
  redshifts for GC formation ($z \geq 15$).
\end{abstract}


\section{Introduction}\label{sec:intro}

Globular Clusters (GCs) are fundamental probes of star formation in
the early Universe and offer a unique local window to the physics and
conditions from the vantage of the evolved Universe.
 GCs harbour one of the
most ancient stellar populations, with estimates for the upper
limit in ages of Milky Way GCs of $13.4\pm 2.2$~Gyr
\citep{Krauss_2003}. Although GCs present a wide range of ages and
metallicities \citep{Peng2006,Harris_2015}, it is interesting to
consider that the older and more metal poor objects could have formed
in the pre-reionization era, and furthermore, that their stars may have 
contributed to the total budget of photons that drove
reionization \citep{Griffen2013,MBK_2017}.

Despite their relevance to star formation, galaxy evolution and
cosmology, there is no clear consensus on the formation of
GCs. 
This stems in part from the large spread in observed GCs properties,
likely suggesting that there is more than one formation
mechanism in play. For instance, the metal rich population of GCs is
believed to be formed mostly in-situ from the inter-stellar-medium
(ISM) of galaxies, with an 
efficiency higher in disks that are more gas rich, 
highly pressurised and turbulent \citep{Kruijssen2015},
conditions that are more prevalent at high redshift, 
and would naturally explain the old ages inferred for
GCs. This scenario is also supported by observations of nearby
interacting galaxies that exhibit distinctive knots of young stellar
associations that could be hypothesised to survive as self-bound GCs
today, as well as studies of young stellar clusters in the MW
and star forming galaxies \citep{Goddard2010,Bastian2013,Adamo2015}. 
In this fashion, metal rich GCs are born out of the ISM in
galaxies that are later re-distributed into a more extended ``halo''
like configuration due to dynamical processes such as scattering by
molecular clouds, mergers and galaxy interactions
\citep{Kravtsov_2005, Prieto_2008}. 

On the other hand, older and more metal poor GCs are consistent with
being accreted from smaller galaxy satellites \citep{vdBergh_2000}
as part of the cosmological assembly of halos. For instance, in the
Milky Way (MW) several GCs have been associated to known stellar
streams and dwarf galaxies \citep{Cohen_2004,Law_2010}. 
Coherent rotation (which could be a an indicator of coherent infall)
has also been observed for M31 \citep{Veljanoski_2014}, a situation
that has also been found in simulations \citep{Veljanoski_2016}.
Another plausible formation
mechanism for GCs has been recently proposed by \citet{Naoz2014}
where non-negligible streaming velocities between the gas and the dark
matter at high redshifts might cause the baryonic collapse of GC-mass
objects outside the virial radius of the dark matter
halos. In a final alternative, metal poor GCs could represent the end product
of galaxy formation on low mass dark matter halos
\citep{Peebles_1984}. This scenario  proposes that GCs form before
reionization {\it at the centres of their own dark matter halos}, in a
similar fashion to dwarf galaxies do, but extending the host halo
mass regime towards lower virial masses, and to baryonic objects
where DM halos have not been detected.
Interestingly, high
resolution hydrodynamical simulations of first galaxies are starting
to support the formation of very compact stellar systems within early
collapsed halos that are structurally similar to the population of GCs
\citep{Kimm_2016,Ricotti2016}.

The idea of GCs inhabiting their own dark halos has been explored in
the past with some attractive conclusions. Three shall be
highlighted: radial distribution of GCs, the scaling between GCs
content and virial mass of the halo and, most importantly, the fact
that GCs will then be predicted to contain large fractions of dark
matter. We discuss each of them individually below. 

\begin{itemize}
\item \emph{Radial distribution of GCs:} Selecting rare peaks of
density fluctuation in the early Universe as sites of GC formation
leads to a present-day distribution of GCs that is highly clustered
around massive host halos such as MW mass and above \citep{Diemand_2005,
  Moore_2006}. Whether the distribution is clustered enough is a
subject of study of the present paper. For instance, the radial
profile of the MWs metal poor GCs has a median radius of $7.1$~kpc
\citealp[(2010 edition)]{Harris_1996} and decays with exponent
$r^{-7/2}$ \citep{Harris_1976}. For comparison, the early work by
\citet{Diemand_2005} populating GCs in high density peaks has given a
consistent radial distribution of GCs around a MW-like halo
along with a relatively compact median radius of $17$~kpc (which is
significantly more extended than the aforementioned blue GCs
population in the MW, although still far from the $\approx 90$~kpc
median radius of the total dark matter halo material).

Furthermore, more recent approaches to this problem have suggested
that these compact distributions of GCs in the MW may be reconciled
with predictions of this scenario by assuming a rather early
reionization time, $z \leq 13$, for the MW region \citep{Bekki_2005,
  Busha_2010}. Although earlier than the suggested reionization time
by the latest Planck analysis, $z_{\rm re} = 8.8^{+1.7}_{-1.4}$,
\citep[][from the Thompson scattering of CMB photons by free
electrons]{Planck_2016}, this may still be accommodated in a
scenario where reionization is patchy \citep[see e.g.][]{Busha_2010,
  Lunnan2012, Griffen2013} and was locally driven earlier either by
the MW itself or by nearby massive structures such as the Virgo cluster
\citep{Iliev2011,Aubert2018}. Encouragingly, the sizes of globular
cluster distributions around external galaxies is beginning to become
available for a large sample of galaxies \citep[Lim et al., in prep.;][]{Georgiev_2010, Hudson_2017, Forbes_2017}, 
providing strong constraints to
this model.

\item \emph{GC abundance that scales with $M_{\rm vir}$:} Due to
the self-similarity of the $\Lambda$ cold dark matter model
($\Lambda$CDM) and its predicted level of substructure \citep{Yang_2011}, a
model where GCs form at high redshift dependent only upon halo mass
sets up a relationship \emph{today} where the total number or mass of
GCs\footnote{The relation with GC number \citep[e.g.][]{Zepf_1993} appears to have more
scatter than using GC stellar mass  \citep{Spitler_2009}.}
is nearly linear in the halo
mass \citep[see also][]{MBK_2017}, with recent estimates of the
normalisation in observations of $M_{\rm GC}/M_{\rm halo}$ of around
$4\times 10^{-5}$ \citet{Hudson_2014}. 
Formation of such a trend from later-time
baryonic processes is more difficult since in general the stellar
distribution or galactic baryon distribution is highly nonlinear in
halo mass \citep[e.g.][]{Moster2013}, causing \citet{Blakeslee_1997}
to memorably comment: ``galaxies do not have too many globular
clusters for their luminosity, they are underluminous for their number
of globular clusters''.
However, it seems that the linear scaling between GC mass and halo mass also arises naturally as consequence of the hierarchical nature of \lcdm, even if they formed through fully baryonic processes, and therefore is a weak probe of GC formation scenarios \citep{ElBadry_2018}.

\item \emph{Dark Matter content in GCs:} The most straightforward
consequence this scenario is the implication that GCs should be surrounded
by an extended and massive dark matter halo, in a similar fashion
as inferred for dwarfs and more massive galaxies. Unfortunately the
baryonic content of GCs is extremely compact with a scale radius of just a 
few parsecs, which in comparison to a dark halo with scale radius of a kpc
\citep{Conroy_2011} implies that within the half light radii we expect the
dark matter content to contribute just a few percent to the matter
density, making the discrimination of its presence or absence a demanding task.
Although stellar dynamics are not available in the dark-matter dominated regions of 
the outer halos,
stellar light profiles and kinematics have been used to attempt a
dynamical modelling able to constrain the matter content of some MW GCs
\citep[e.g. ][]{Lane_2010, Conroy_2011, Ibata_2013, Penarrubia_2017}
which have not found evidence for DM halos, albeit with the additional 
invocation of assumptions about levels of equilibrium/relaxation and tidal heating and stripping.
Interestingly, high mass-to-light ratios have been
detected in GCs outside the MW \citep{Taylor2015}, although the distribution seem
so compact that is better explained by central black holes
\citep{Bovill_2016} or some peculiar stellar population (T. Puzzia,
private communication) than by a surrounding dark mater halo.
\end{itemize}

Further interesting predictions from this simple model
includes the dynamics of GCs (now dictated by cosmological infall of
subhalos) and the existence of very few GCs expected to have survived
until $z=0$ in isolation, i.e. outside the virial radius of galaxies and
clusters. Whereas promising new measurements from observations are
starting to constrain the kinematics of GCs
\citep{Zhu_2014,Deason_2013, Napolitano_2014, Spitler_2012}, the
searches for `intergalactic' GCs in the large SDSS data-set has
not confirmed any candidates so far \citep[][]{diTullio_2015,
  Mackey_2016}.  
Therefore, as observational data starts to constrain the full space
of predictions by the model, the jury is still out on whether primordial GCs
could explain (or partially explain) the population of metal poor GCs
around galaxies.

Studying the formation of GCs in cosmological hydro simulations is
extremely challenging given the large dynamical range needed to
resolve the inner structure of GCs (pc-scale) together with the
formation and environment of halos on Mpc-scales
\citep{Kravtsov_2005, Ricotti_2016, Mistani2016, Li_2017,
  Kimm_2016,Renaud_2017, Pfeffer2018}. We instead employ a
complementary approach where N-body only cosmological simulations are
used to identify the possible sites of formation of GCs and a
`tagging' criteria to follow the GC evolution (in a collisionless way)
up to redshift $z=0$. We explore the implications of GCs forming as
the central objects in dark matter halos at the redshift of
reionization with specific focus on the conditions required to satisfy
observational constraints on abundance, radial distribution and the
observed (or lack thereof) intergalactic globular cluster (IGC)
distribution. We expand previous studies using a similar technique, by
looking at the GC population predicted in a wide range of halo mass
($M_{200}$ from $10^{11}$-$10^{12.6}$~\Msun), instead of analysing the
MW-like halos only. Finally we also consider the implication w.r.t. DM
in the GCs \emph{today}, and to what extent observations of the
extended stellar halos (of GCs) can constrain this.

This paper is organised as follows. In Section~\ref{sec:sims} we
describe our dark matter only numerical simulation that is large
enough to capture several MW-like objects, yet high enough resolution
to identify the progenitor halos that could be hosting GCs today, and
how we tag our GCs. In Section~\ref{sec:gc_pop} we make a comparison
of this population to the observed GC population, with particular
attention to the abundance of GCs and their radial distribution. 
In Section~\ref{sec:dynamics} we follow their orbital
dynamics around their hosts and in Section~\ref{sec:IGCs} consider the
abundance of external `intergalactic' GCs. In
In Section~\ref{sec:discussion} we summarise and
conclude.

\section{Cosmological simulation}\label{sec:sims}
In order to identify dark matter halos in \lcdm\ that are numerous
enough at redshift 10 to be the hosts for an old GC population one
needs to resolve halos down to a mass of a few $\times 10^7$~\Msun,
and yet be large enough to contain several MW like objects
(i.e. within a factor of a few of a halo mass of $10^{12}$~\Msun)
implying one needs to simulate a volume of several hundreds of
comoving Mpc$^3$. Analytic models such as \citet{PS_1974} can be used
if one is only interested in the mass evolution and assembly history
(e.g. as done for a similar GC formation scenario by
\citealp{MBK_2017, ElBadry_2018}), yet if one needs the dynamics and the geometry
one must turn to N-body simulation.

As such we choose a box of $10$~Mpc comoving on each size, uniformly
resolved with $512^3$ particles with mass of $2.95\times 10^5$~\Msun\
each (i.e. $\approx 340$ particles for a $10^8$~\Msun\ halo). We
created initial conditions using our own code
\href{https://github.com/pec27/lizard}{Lizard} (Creasey et al., in
prep.) with the Planck cosmology with present day mass fraction
$\Omega_{\rm M}=0.315$, dark energy $\Omega_\Lambda=0.685$, Hubble
constant $H_0 = 67.3\;\rm km\, s^{-1}\, Mpc^{-1}$, normalisation
$\sigma_8=0.829$ and spectral index $n_{\rm s} =0.9603$, realised at a
redshift of $127$.

The gravitational collapse of this dark matter distribution was
evaluated with the N-body code {\sc Arepo} \citep{Springel_2010},
using a softening length of $490$~pc physical or $1.95$~kpc comoving,
whichever is the smaller, and collapsed structures of more than 32
particles identified with {\sc Subfind}.

\subsection{Tagging of GCs}\label{sec:tagging}

In order to infer the orbits of a mass component that has formed in
the centres of these early-collapsed halos we tag the most bound particle for all
groups with virial mass $M_{200}^{\rm tag} \geq 10^8$~\Msun\ at a
redshift of $z_{\rm tag} =8.65$, the mass cut chosen to match the
number density expected for the MW (later we consider higher and lower
mass cuts to examine the effect of this choice). The redshift for
  tagging in our fiducial model should be interpreted
  as the average redshift of reionization (a later assembly being 
assumed unable to form a GC) and has been chosen to be
  consistent with Planck optical depth constraints \citep{Planck_2016}. 
We follow
these to redshift zero where the majority of these particles can still
be associated with bound structures, however we do not discard those
that do not since even if one does not resolve the structure, it is
still expected to exist \citep{vdBosch_2018} and the most bound
particle is used as our best tracer for dynamical purposes.

The absent baryonic component of these simulations would (if present)
collapse into tightly bound objects (galaxies) with scale lengths of
$O(1)$~kpc (e.g. \citealp{Shen_2003}) at the centres of these
halos. These will affect the GC population both due to baryonic
contraction of the orbits and tidal disruption of those GCs for whom
the pericentric distances become comparable to the galactic scale
length. The quantification of this we defer until
Section~\ref{sec:dynamics}.

\begin{figure}
\includegraphics[width=\columnwidth]{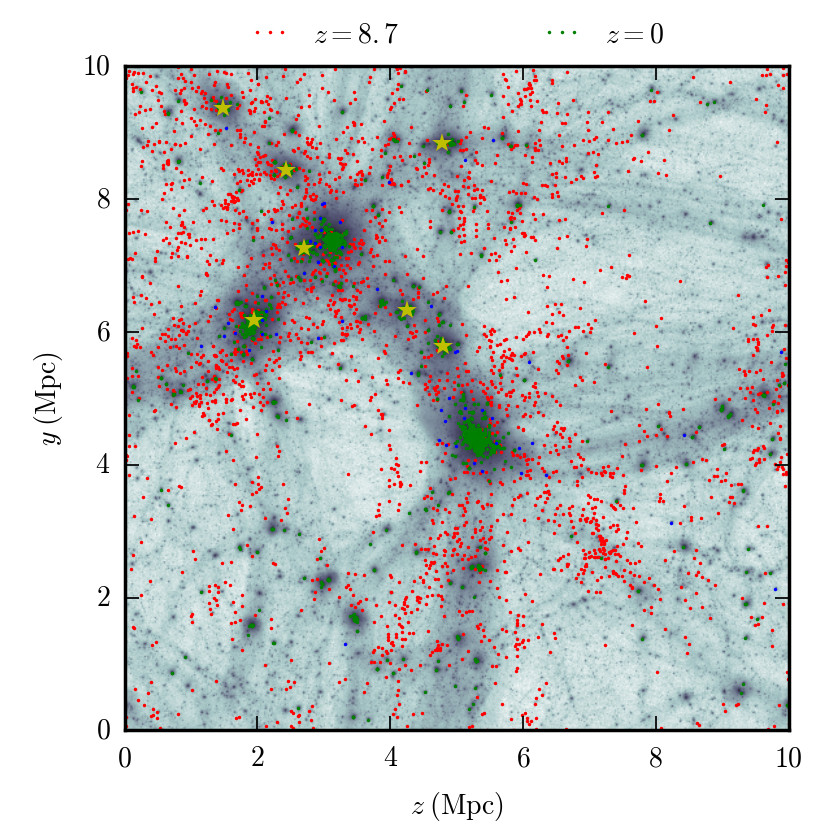}
\caption{Projection of GC candidate positions identified at redshift
  $8.7$ (\emph{red dots}) as halos of $M_{200} > 10^8 \; \rm M_\odot$
  and tracking the most bound particle to redshift $0$ (\emph{green
    dots}), with \emph{blue dots} denoting structures that are still
  the most massive in their host halos. 7 MW-like halos (see text) 
  have $z=0$ positions denoted by \emph{yellow stars}, and 
  \emph{background shading} represents dark matter
  density at $z=0$.}
\label{fig:projection}
\end{figure}

For an overview of our cosmological simulation we show in
Fig.~\ref{fig:projection} a projection of the present day DM density,
7 selected MW-mass halos along with initial ($z=8.65$) and final
($z=0$) positions of the tagged GCs in red and green
respectively. Immediately apparent is both the strong clustering of
the initial GC sites (i.e. these are relatively extreme objects at
formation) and the corresponding clustering of their final positions
near the centres of massive halos today.

\section{The GC population}\label{sec:gc_pop}
In this section we compare the distribution and dynamics of the halos
identified in Sec.~\ref{sec:sims} with the observations of the
globular cluster population. We begin in Sec.~\ref{sec:abundance}
by examining the number density of GCs per host and then in 
Sec.~\ref{sec:spatial} their spatial distribution within their host
halos. When possible we compare only to the low
metallicity/blue population. We work under the assumption that the
metal rich and red population of GCs is formed in-situ out of baryonic
processes in the central galaxy and not well modelled by the hypothesis
of GCs formed in their own dark matter halos.

\subsection{GC abundance} \label{sec:abundance}

\begin{figure}
\includegraphics[width=\columnwidth]{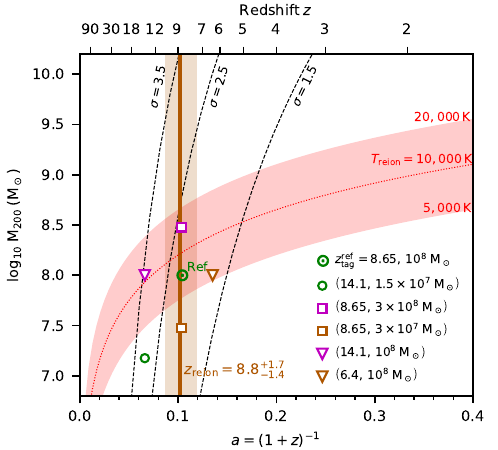}
\caption{The tagging times (expansion factors) and masses used in this work, in comparison
  with other constraints.
  \emph{Dotted green circle} indicates the fiducial $z_{\rm tag}=8.65$, $M_{\rm tag}=10^8$~\Msun,
  whilst \emph{open coloured symbols} indicate other taggings (see values inset), with colours corresponding to Fig.~\ref{fig:gc_vs_mvir}. 
  \emph{Brown vertical line} and \emph{shaded} indicate the Planck~2016 reionization time
  estimate and uncertainty. 
  The \emph{red line} indicates halo masses that correspond to a reionization temperature of $10^4$~K, 
  with the effect of raising or lowering to $20,000$~K or $5,000$~K given by the \emph{red shaded region}.
  \emph{Labelled black lines} indicate the rarity of these halos (the number of sigma of the expected linear overdensity from the mean, see text).
  The models in \emph{green} approximately match the number of blue GCs for a $10^{12}$~\Msun halo (see Fig.~\ref{fig:gc_vs_mvir}).}
\label{fig:a_vs_m}
\end{figure}

\begin{figure*}
  \includegraphics[width=2\columnwidth]{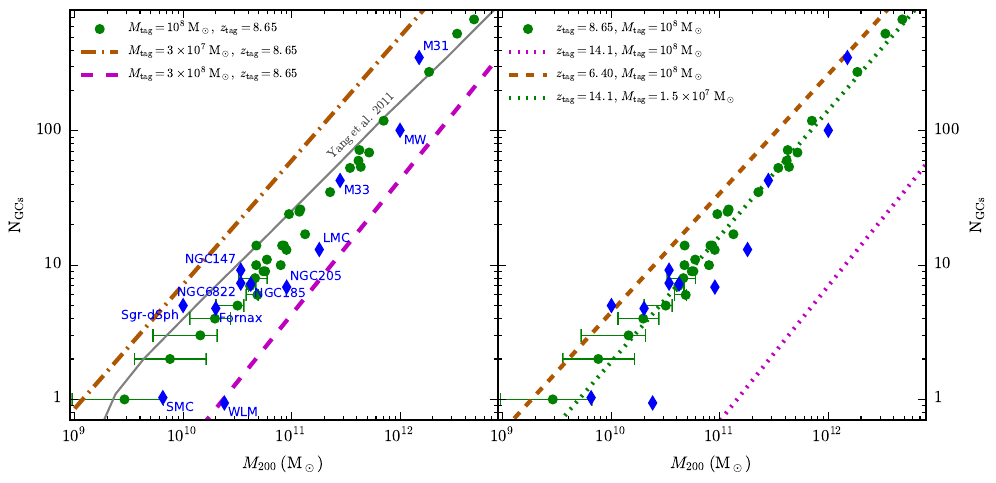}
  \caption{GC counts as a function of halo mass for our simulation
    (systems with more than 8 GCs are shown as individual \emph{green points}, 
    otherwise we indicate the range with error bars for the
    15 and 85th percentiles in mass).  \emph{Blue diamonds} indicate
    the number of inferred blue GCs of local group galaxies (with halo
    masses inferred from the abundance matching of \citealp{Moster2013})
    and the fraction of blue GCs using the conversion in Eqn.~(\ref{eq:harris_blue})
    for all except the MW (where the discrimination is made with
    Fe/H).  
    \emph{Left panel} shows the expected substructure scaling \citep[from
    ][]{Yang_2011} in \emph{grey line}.
    \emph{Coloured dashed lines} indicate the relations for our simulation 
    with different tagging masses, \emph{dashed magenta line} is for $3\times 10^8$~\Msun\ and 
    \emph{dot dashed brown line} is for $3\times 10^7$~\Msun.
    In the \emph{right panel} we show the effect of different redshift cuts, with
    $z_{\rm tag}=14.1$ in \emph{magenta dotted} and $z_{\rm tag}=6.4$ in 
    \emph{brown dashed line}. \emph{Green dotted} line shows a combination 
      of a lower mass cut of $M_{\rm tag}=1.5\times 10^7$~\Msun\ at the higher 
      tagging redshift ($z_{\rm tag}=14.1$) that has been chosen to match the current blue GC abundances.}
\label{fig:gc_vs_mvir}
\end{figure*}

The most basic question to ask of a GC formation model is whether it
predicts the right abundance of GCs as a function of their hosts.  The
tightest correlation is believed to be between total mass in
globular clusters vs. halo mass \citep{Spitler_2009}, although
one can use GC number as a proxy for GC mass as well \citep[e.g. ][]{Blakeslee_1997}.
Using number of GCs instead of mass eases the comparison with our
simulations since we have not followed the detailed baryonic processes
(i.e. environment dependent formation and subsequent stripping) which
would allow us to assign masses per GC.

As explained in Sec.~\ref{sec:tagging}, we use our fiducial cut to
seed all halos above $M_{\rm vir} = 10^8$~\Msun\ at $z=8.65$ with a
GC. We then follow the most bound particle in these halos at $z=0$ to
predict the position and velocities of those GCs at present day. As
shown in Fig.~\ref{fig:projection}, due to hierarchical assembly,
painted GCs are today highly clustered around more massive halos, in
good agreement with the distribution of observed GCs. 
We emphasise that this particular choice of tagging parameters is by no means unique; there is a degeneracy between mass and redshift for tagging that can reproduce a similar abundance of subhalos at z=0 \citep[see also][]{Diemand_2005}. 
In Fig.~\ref{fig:a_vs_m} we illustrate this degeneracy in terms of the rarity of density peaks given a combination of mass and redshift for tagging. We also include expectations from reionization, in particular,
our fiducial
tagging corresponds to halos above a reionization temperature of just below $10,000$~K at a time consistent with the Planck measurements, although this correspondence is not exact since reionization is expected to be neither instantaneous nor entirely homogeneous. For this reason we consider 5 other combinations of time and mass, indicated with different symbols. 
A previous estimate by \citet{Moore_2006} of the rarity of material needed to forms globular clusters corresponded to halos whose (linear) over-densities are approximately $2.5\sigma$ above the mean. For comparison we have included the halo mass corresponding to $1.5$, $2.5$ and $3.5\sigma$ (using a sharp-$k$ filter) as a function of expansion factor.

In Fig.~\ref{fig:gc_vs_mvir}, we convert these taggings into 
the number of GCs within the virial radius (defined here as the radius that encloses $200\times$ the critical density) of all
identified halos in our simulation output at $z=0$ and plot the resulting
abundance of GCs per halo, where we have only included halos with at least 1~GC.
This is shown as green solid symbols, using individual dots for
halos with more than 8 GCs and otherwise indicating the range with
error bars for the 15 and 85th percentiles in mass. This is compared
to GC observations in the local group compiled in \citet{Mackey_2015},
where we have converted V-band magnitude to stellar masses for each host
galaxy assuming a mass-to-light ratio of $1$  and estimated halo masses
based on the abundance matching prescription from
\citet{Moster2013}. Additionally, local group values (all except for
the Milky Way) have been corrected by the estimated fraction of blue
GCs using:
\begin{equation}\label{eq:harris_blue}
f_{\rm blue} = \min\left( \frac{M_{200}}{10^7 \; {\rm M_\odot}}, 1\right)^{-0.07}
\end{equation}
taken from \cite{Harris_2015}. For the MW, we have applied instead a
metallicity cut, including all GCs with Fe/H$<-1$ as blue \citep[2010 edition]{Harris_1996}.

The left panel of Fig.~\ref{fig:gc_vs_mvir} show a reasonable good
agreement between local observations and the predictions of our
model. We highlight that while the agreement on the \emph{normalisation} 
is obtained by construction-- our fiducial mass cut
$10^8$~\Msun\ was chosen to reproduce approximately the abundance of
blue GCs in the MW, see Sec.~\ref{sec:tagging} -- the
\emph{scaling}, which is almost linear in halo mass, emerges naturally
within $\Lambda$CDM since the number of high redshift progenitors
scales approximately linearly with the halo mass. 
The grey solid line on the left panel of Fig.~\ref{fig:gc_vs_mvir}
corresponds to
the scaling of the number subhalos of $10^8$~\Msun\ accreted by halos
as a function of mass \citep[taken from][]{Yang_2011}.
Further
discussion of the application of this approach to GCs can be found in
\citet{MBK_2017} and \citet{ElBadry_2018}. One statistic not shown in this plot is the fraction of halos which
have no GCs. For our fiducial mass cut, around 50\% of galaxies in an $M_{200}\approx 5\times 10^9$~\Msun\  
halo (with stellar mass $M_\star \approx 10^6$~\Msun) have no GCs, falling sharply with increasing halo mass.

Fig.~\ref{fig:gc_vs_mvir} also explores the effects of assuming a
different mass cut (left panel) or redshift cut (right panel) for the
tagging of GCs. In agreement with previous work
\citep[e.g. ][]{Diemand_2005}, more restrictive cuts in terms of
density fluctuations at high redshift would result in a lower number
of primordial GCs identified around more massive halos today. For
example, a higher mass cut $M_{\rm vir} > 3\times 10^8$~\Msun\ to seed GCs (magenta
dashed, left panel) or an earlier reionization time
$z_{\rm tag} = 14.1$ (magenta dotted line, right panel) is
reflected on a smaller fraction of halos qualifying as GCs formation
sites and therefore a smaller abundance of GCs per halo at $z=0$
compared to our fiducial model. On the other hand, tagging less
massive halos (brown dot-dashed line, left panel) or our fiducial mass cut
but at a later reionization time $z_{\rm tag}=6.4$ (brown dashed line, 
right panel) would both result on a higher GC occupation per
halo today. As shown in the next section,
these variations on the abundance of GCs given how ``rare'' the
GC-hosting density peaks are at high redshifts correlate strongly with
their expected clustering at $z=0$, which can be used in combination
with observations to better constrain this scenario.

\subsection{GC spatial distribution}\label{sec:spatial}

Our next comparison is the predicted radial distribution of GCs around
massive hosts at the present day. For reference, the number density of
globular clusters in the Galactic halo is observed to decline steeply
with distance to the MW, with $n(r)\propto r^{-3.5}$ as average best
fit relation \citep[e.g. ][]{Harris_1976,Helmi_2008}. This is
comparable to the radial distribution of halo stars in numerical
simulations of galaxies but is significantly more concentrated than
the expected distribution of dark matter or even dwarf satellite
galaxies \citep[e.g. ][]{Abadi2006}. We have selected the seven halos
in the `MW-like' halo mass range of $M_{200} = 10^{12 \pm 0.5}$~\Msun\
and checked that the radial distribution obtained is consistent with
this value, a fact that has previously been highlighted
\citep{Diemand_2005} as an appealing feature of this mechanism for GC
creation (and similarly for halo stars). 

\begin{figure}
\includegraphics[width=\columnwidth]{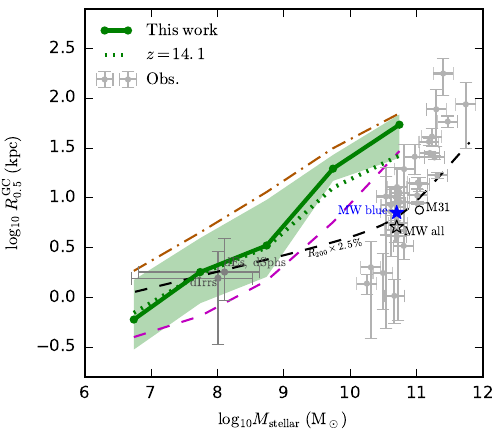}
\caption{Median radii of GCs from their hosts. \emph{Green line} shows
  the median relation for our simulation, with \emph{shading} for 15
  and 85th percentiles.  
  \emph{Open black star} and \emph{circle} are for the MW and
  M31 respectively, \emph{filled blue star} is the blue-only GC
  population for the MW.  Observed radii are \emph{grey points} taken
  from \citet{Hudson_2017} in addition to dIrrs (\emph{orange
    outline}) and dEs and dSphs (\emph{red outline}) from
  \citet{Georgiev_2009}.  The black star is the 17~kpc median radius
  quoted in \citet{Diemand_2005}.  The effect on populations of higher
  and lower mass cuts (as for Fig.~\ref{fig:gc_vs_mvir}) is shown in
  \emph{magenta dashed} and \emph{brown dot-dashed lines} respectively, and
    the \emph{green dotted line} represents the $z=14.1$ cut.
  \emph{Dashed black line} indicates $2.5\%$ of the virial radius, approximately
  matching the MW median GC radius.}
\label{fig:halfrad}
\end{figure}

However, the galactic GC distribution seems to flatten for the
interior $50\%$ of GCs and possibly steepens for the outermost (see
e.g. Fig.~7 of \citealp{Harris_1976}) making the radial profile of GCs
somewhat radius dependent and also well constrained only for the MW
and a short handful of other galaxies. An alternative metric to
quantify the radial distribution of GCs about their hosts is to use the
median distance, i.e. the radius enclosing half
the number of GCs per host, which has been recently observationally constrained
on hundreds of galaxies over a wide mass range
\citep{Georgiev_2009,Hudson_2017}. In
Fig.~\ref{fig:halfrad} we show the median half-number size,
$R_{0.5}^{\rm GC}$, predicted by our fiducial model as a function of
the stellar mass of the host galaxy at $z=0$ (green solid line,
stellar mass computed assuming abundance matching model by
\citealt{Moster2013}). Green shaded region indicates the
$15\%$-$85\%$ quartiles of the distribution. For our fiducial model,
the half number radius of the GC system is comparable to $\sim 14\%$
of the virial radius of the host halos, whilst for the MW it is only about $2.5\%$,
a fraction which we plot for other halo masses for comparison.

The predicted size for the GC distribution can vary with the
assumptions made to tag these GCs. For instance,
Fig.~\ref{fig:halfrad} shows the median half-number radius expected if
the mass cut for tagging GCs at $z_{\rm tag}=8.6$ is changed to a
smaller $M_{200}>3\times 10^7$ or larger $3\times 10^8$\Msun\ compared to our
fiducial value. As expected, when the tagged halos are more massive,
or correspondingly ``rarer'' density peaks at $z_{\rm tag}$, the GCs
at $z=0$ are more clustered (dashed magenta line) than in our
fiducial model (solid green). On the other hand, a lower mass cut
results in a more extended distribution (dot-dashed brown line). Similar
behaviour is observed for a higher or lower redshift cuts at fixed
halo mass. These variations are interesting when combined
with predictions for the number of GCs expected in each case. As
discussed in Fig.~\ref{fig:gc_vs_mvir}, the abundance of GCs within
halos at $z=0$ also depends on the assumptions for tagging, with
larger abundances associated to less extreme cuts to host a GC in the
early Universe (either lower mass or lower redshift). {\it This
  highlights a strong prediction of this model: a fundamental relation
  between the size of the GCs distribution and the abundance of GCs
  per host halo at present day}.

We compare the GC distribution in our models to observations in
Fig.~\ref{fig:halfrad}, showing with symbols plus error-bars available
measurements on the half-number radius of GCs on external galaxies
\citep[data taken from
][]{Georgiev_2009,Hudson_2017}\footnote{\citet{Hudson_2017} reports
  the $R_{\rm e}$ of GC systems, where $R_{\rm e}$ is the half light
  radius of the best fit de Vaucouleurs profile to the GCs
  distribution. \citet{Georgiev_2009} results are multiplied by a
  factor~$4/\pi$ to convert the median projected radii to 3D}.  Data
for the blue GC population of the MW (blue star) and M31 (open circle)
are taken from in the case of the MW and from \citealp[(2010
edition)]{Harris_1996} and for M31 from \citet{Galleti_2004} complemented with outer
halo clusters from the Pan-Andromeda Archaeological Survey \citep{Huxor_2014}. The result of our
fiducial model (chosen to roughly reproduce the abundance of GCs
today) is consistent with observations of low mass galaxies
($M_{\rm stellar}~10^8$~\Msun) but seems to overestimate the sizes of
the GC distributions for larger masses. For instance, for stellar
masses in the range $10^{10}$-$10^{11}$~\Msun\ the half number radius
predicted is $54$~kpc ($25$-$69$~kpc for 15 \& 85th percentiles
respectively) which should be compared to the average $\sim 10$~kpc
expected for MW-like galaxies. We have checked that including the
effects of baryonic contraction in the central potential of these
halos or changing the weight of the tagging technique can shrink the
expected sizes by around 6~kpc, but is not enough to bring the results
into agreement (see Appendix~\ref{sec:tests} for details).

Instead, adopting a higher threshold mass for tagging (magenta curve) 
helps accommodate the half number radius with observations of
the MW, at the expense of explaining the number counts 
(see Fig.~\ref{fig:gc_vs_mvir}). Raising the reionization time 
to $z_{\rm tag}=14.1$ and simultaneously reducing to the tagging mass
to fit the number counts (see Fig.~\ref{fig:gc_vs_mvir} green dotted line) 
also improves the radial distribution, although is still in significant 
tension with the observations. 
We have calculated models where an infalling
distribution is combined with a compact distribution produced in some galactic
process, and find that in order to match the median radii at most 60\% of the 
GCs could come from an infalling population.

At the other extreme, if the GC progenitors were tagged randomly with dark matter particles, their median radii would match that of the dark matter (not shown), at around 90~kpc for the MW and much too extended compared to the observations, even though it would be possible to match the abundances (as shown by \citealp{ElBadry_2018}).

We note that we have ignored the process of tidal disruption of GCs,
that may impact around half of the GCs as they orbit the main halo
(e.g. \citealp{Prieto_2008, Muratov_2010, Choksi_2018}, though \citealp{Carlberg_2018} have found values nearer unity), 
where GCs with smaller apocentric radii are preferentially disrupted into streams. 
Such a preferential removal of GCs would only exacerbate the tension with the observed distribution.

The compact
distribution of GCs around the MW has been highlighted before as a
hint for an earlier reionization time in our Galaxy \citep[$z\sim 15$,
see ][]{Bekki_2005}, which could be a sign of patchy reionization. In
fact numerical simulations have shown that larger halos and clusters
are believed to reionize earlier due to the UV emission from their
progenitors \citep{Alvarez_2009, Li_2014} and partial evidence
for this is also found in observations \citep{Spitler_2012}. A higher
than average redshift of reionization for the MW seems preferred not
only from their GC distribution but would also help alleviate
the missing satellite problem in the Local Group \citep{Busha_2010,
  Li_2014}. However, we notice that the MW or even M31 in
Fig.~\ref{fig:halfrad} are not far from the average of external
galaxies with the same mass, and therefore GC formation models should
be able to account for this rather compact GC distribution without
invoking particular outliers in the reionization history. Based on our
results, current measurements for half number radius of metal poor GCs
suggest that about half and up to $60\%$ of GCs in MW-like objects
could have been formed in early-collapsed halos at high redshift. In what
follows, we study their expected kinematics.

\section{Orbits and dynamics of GCs}\label{sec:dynamics}

GCs that formed at the centres of their own DM halos will follow 
orbits and kinematics predicted primarily by the $\Lambda$CDM
scenario. We use our simulations and tagging technique to explore the
present-day orbits of GCs around MW-mass hosts. Unless explicitly
mentioned, we focus on our fiducial tagging model, although the
results presented do not depend significantly on this assumption. 

\begin{figure}
\includegraphics[width=\columnwidth]{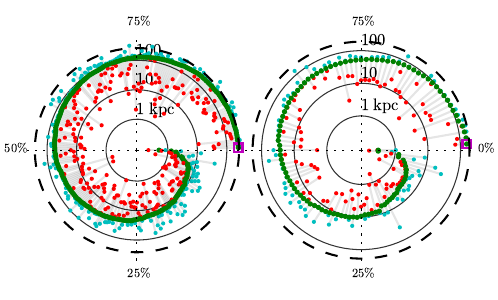}
\caption{Pericentre, apocentre and current radius for the tagged GCs
  in the most massive halo (\emph{left panel}) and 2nd-most massive
  (\emph{right panel}) from Fig.~\ref{fig:projection}. 
  \emph{Red points} indicate the pericentric passage and \emph{cyan points}
  the apocentre (calculated as in Sec~\ref{sec:dynamics}). Orbits are
  ordered in azimuth by their current radius (\emph{green points}, 
  starting from \emph{magenta square}), such that the azimuthal
  direction is the quantile, with percentiles marked with dotted lines.
  Labelled \emph{solid lines} indicate radial distances in
  kpc (spaced logarithmically), whilst the \emph{dashed black line}
  indicates $R_{200}$. }
\label{fig:orbits}
\end{figure}

Dark matter halos within $\Lambda$CDM have radially biased orbits
\citep{Wojtak2009, Wojtak2013} with apocentres that are
significantly larger than the pericentres. Numerical simulations
have shown that subhalos (and therefore satellite galaxies inhabiting
those subhalos) inherit such elliptical orbits
\citep{Sales2007,Iannuzzi2012}, an effect that contributes to the
disruption and lowering of star formation in satellites, even for
those seemingly today in the outskirts of groups and clusters. If blue
GCs --or a fraction of them-- are expected to form within DM halos, we
might expect their orbits to have an ellipticity comparable to
satellites, with present-day positions correlating poorly with their
minimum distance to their hosts.

In order to characterise the orbits of our GC candidates we need a
time sampling that is significantly higher than the outputs of the
snapshots in the simulation. A more accurate orbit parameter
estimation can be achieved by taking the present day position and
velocity of GCs and integrating their orbits in an analytical
potential. We approximate the host potentials as NFW profiles
\citep{Navarro_1997}.  The two NFW parameters are the $M_{200}$ and
the scale radius $r_{\rm s}$ (determined from the relation $r_{\rm
  s}\approx R^{\rm vel}_{\rm max}/2.16$, where $R^{\rm vel}_{\rm max}$
is the radius of maximum circular velocity in the spherically averaged
mass profile). The pericentric and apocentric orbital passages of the
GC candidates are then calculated from the radial and tangential
velocity components $v_{r}$ and $v_\perp$ relative to the halo
centres, along with their current radius.

As an example, we show in Fig. ~\ref{fig:orbits} the relation between
present day distance (green), pericentre (red) and apocentres (cyan)
of tagged GCs for the two most massive MW-like halos in our
simulation.  In this polar representation, GCs are sorted according to
their present day distance, moving counter-clockwise from the furthest
to the nearest and starting on the right magenta square. 
Solid thin lines indicate lines of constant radius at 1, 10
and 100 kpc from the host, while the dashed circle indicates the
virial radius of the system 260 and 189~kpc for left and right,
respectively). A circular orbit in this representation would have
three distances coinciding on the same point. Instead, an elliptical
orbit has a large separation between the corresponding red and cyan
dots, with the green symbol at some intermediate distance according to the
phase of the orbit.

Fig.~\ref{fig:orbits} highlights that, as expected, GCs formed with
this cosmological origin are characterised by quite elliptical orbits,
with apocentres that are on average (median) a factor $\sim 4$ farther than
their pericentres. However variations can be large, with a significant fraction
of GCs that today are found near $\sim 100$ kpc that could have
ventured in the past within just $10$ kpc of their
hosts. Interestingly, this figure also highlights an important
prediction: very few (although non zero) number of GCs have apocentric
distances that go beyond the $R_{200}$ of the system and may be
interpreted in observations as ``intra-galactic'' objects when they
are actually bound. We return to quantifying this in
Sec. ~\ref{sec:IGCs}.

\begin{figure}
\includegraphics[width=\columnwidth]{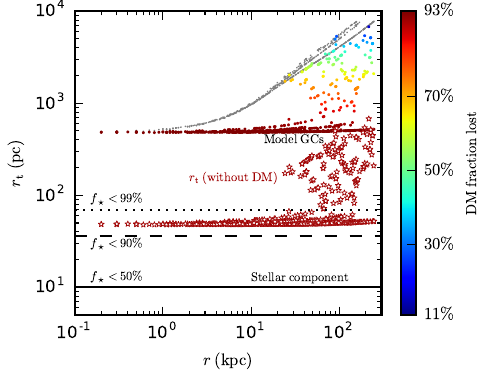}
\caption{ Estimate of the tidal truncation radius from their pericentric passages, 
  with and without the DM halo,
  for the GCs around the 7 model MWs from Fig.~\ref{fig:projection}.
 \emph{Coloured points} indicate the tidal radius using the DM satellite mass
 in Eqn.~(\ref{eq:r_t}), where the colours correspond to the fraction of DM stripped (from 11-93\%).
 \emph{Red stars} indicate the tidal radius using only the stellar component (i.e.
 for those where the DM halo has been stripped). For comparison, \emph{horizontal black solid, dashed} and \emph{dotted lines} 
 indicate a fiducial GC stellar distribution, as the radii that contain $50$, $90$ and $99\%$ of the stellar fraction. }
\label{fig:truncation}
\end{figure}

The large ellipticity in GCs orbits means that they could have been in
the past exposed to much stronger tidal forces than estimated for
their present day position, with important consequences on the GCs
tidal disruption and, especially, of their surrounding dark matter
subhalo. Indeed, Fig.~\ref{fig:orbits} suggests that the dark matter
content of GCs shall not be judged as intact even for GCs that {\it
  today} are seen in the outskirts of their hosts.  How much of their
initial dark matter mass might still be associated to blue GCs is
therefore dependent not only on their current distance but also of
their past orbital history. For instance, at our resolution, only
$20\%$ of the GC-hosting subhalos whose most bound particle is around
$100$~kpc today has retained a resolved DM-substructure that is
identified by our subhalo finder (32 particles or more). This number,
although indicative of the overall trend, is certainly strongly
dependent on the numerical resolution of our simulation \citep[see
e.g.][]{vdBosch_2018}. Instead, we estimate analytically
the expected truncation radius and retained dark matter mass fraction
for the subhalos in Fig.~\ref{fig:truncation}.

To be conservative, we estimate the tidal radius and retained mass
fraction at the pericentric distance, when the
background potential is maximal and the effects of tidal disruption
are strongest. Taking the best-fit NFW profile to our MW-like host halos, 
we estimate the tidal radius $r_{\rm t}$ of our tagged subhalos from
\begin{equation}\label{eq:r_t}
\left( \frac{r_{\rm t}}{r} \right)^3 = \frac{M_{\rm sat}}{\left(2- \frac{\de \ln M_{\rm host}}{\de \ln r}\right) M_{\rm host}(<r)} 
\end{equation}
with $M_{\rm host}$ and $M_{\rm sat}$ the total mass of the
host and satellite dark matter halo \citep[see][]{Springel08}, and $r$ being either the pericentric distance to the host or the current distance (see Fig.~\ref{fig:truncation}). We
also account for the effects of baryons in the central halo by adding
to the potential a disk of mass $10^{11}$~\Msun\ and scale
radius $3$~kpc.  As shown in
Fig.~\ref{fig:truncation}, the tidal field of the MW generally only 
strips the subhalos down to $>500$~pc, well above the
observed half-light radius of the stars in GCs ($r_h \lesssim 10$~pc,
see solid black line), even for orbits with small pericentric
passages. This is somewhat misleading, however, since 
many of these will have lost the majority of the dark matter
(these tidal radii correspond to $11$-93\% of their dark matter).
If the DM stripping becomes non-linear (e.g. for multiple pericentric passages),
then the relevant satellite mass in Eqn.~(\ref{eq:r_t}) is only the stellar mass, 
and we additionally plot the calculation for these in Fig.~\ref{fig:truncation} (star symbols).
For comparison we also show enclosed mass fractions for a fiducial \citet{King_1966} profile with half mass radius of 10~pc and concentration $1.5$,
and the overlap of a significant fraction suggests that many of these would have appreciable tidal tail, such as those seen in \citet{Carlberg_2017}.

Whether or not the DM content of the relatively undisturbed 
subhalos should still be observable is the subject of study of a
forthcoming paper (Creasey et al., {\it in prep}), but we highlight
here that one of the strongest predictions of this GC formation model,
namely the existence of a massive and extended dark matter halo
surrounding each GC, should be evaluated with care due to the highly
radial orbital structure expected in this scenario for GCs around
their host galaxies added to the the very compact distribution of
stars mapping too little of the expected extension of this dark matter
component.

\begin{figure}
\includegraphics[width=\columnwidth]{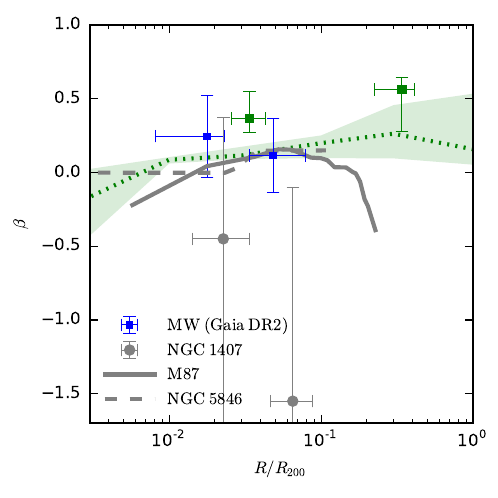}
\caption{Velocity anisotropy ($\beta$) as a function of distance
  normalised to the virial radius for the GCs in the 7 MW mass halos
  from our simulations (\emph{green square symbols}). Due to low number
  statistics we show only 2 radial bins.  Error-bars correspond to the
  15-85\% percentiles. GCs in this scenario are expected to be on
  strongly biased radial orbits ($\beta \geq 0.3$) which are slightly
  more radial than the underlying dark matter particles in the host
  halo (\emph{green dotted line} and \emph{shaded region} showing 
the 15-85th percentiles of the distribution).  We compare with a few available observations from
  the literature: GC anisotropies for the blue GCs of the MW are shown in \emph{blue squares} with errors, for M87 are shown as the \emph{solid grey line} 
  \citep[from][Fig.~12]{Zhu_2014}, for NGC~5846 in
  \emph{dashed grey} \citep[from][Fig.~5]{Napolitano_2014}, and for
  NGC~1407 in \emph{grey circles with error bars}
  \citep[from][Fig.~3]{Spitler_2012}.}
\label{fig:beta}
\end{figure}

An alternative way to quantify the ellipticity of the orbits is through the
anisotropy parameter $\beta$ defined as: 
\begin{equation}\label{eq:beta}
\beta = 1 - \frac{1}{2} \frac{\left< v_\perp^2\right>}{\left< v_\parallel^2 \right>} \, ,
\end{equation}
where $\left< v_\perp^2\right>$ and $\left< v_\parallel^2 \right>$ 
are the velocity dispersion in the radial and tangential
directions. 
Even in a simulation where there is no uncertainties on the tracer velocities this generally requires large sample sizes (since both quantities need to be estimated),
and so
we compute the average $\beta$ profile of
the tagged GCs as a function of distance (normalised to the virial
radius) for the stacked sample of our 7 MW-mass halos in two radial bins 
(median and 15-85\% percentiles shown in green square symbols with error-bars),
which we choose located at $0.03$ and $0.3$ $R_{200}$. As
expected from Fig.~\ref{fig:orbits} the predicted orbital structure
for GCs is strongly biased radially ($\beta > 0$), in particular at
large radii. Notice that the anisotropy of the orbits of GCs traces
roughly that one of the underlying dark matter halo of the host (see
green dotted and shaded regions), although with slightly larger
radial component than the dark matter particles \citep[in agreement
with ][]{Diemand_2005}.

Observationally, measuring the anisotropy of GCs is an even more challenging task,
however some observational constraints are becoming
available. For the case of the MW, \citet{GaiaGCs_2018} provide
proper motions for 75 GCs, and we show the velocity anisotropy estimated 
for the blue GCs in two radial bins in Fig.~\ref{fig:beta}. 
Although not strongly constraining (primarily due to the low number statistics), they do suggest a small positive radial anisotropy consistent with a radial infall model.

Beyond the MW, for other external $\sim L_*$
galaxies (e.g. M31) one can apply spherical Jeans modelling to infer
the missing components\footnote{In the case of M31 the GC spatial and
  velocity correlations are visually apparent \citep{Djorgovski_1997},
  previously \citet{Veljanoski_2014} assumed an isotropic ($\beta=0$)
  velocity dispersion in the 30-100~kpc range.}, but the
mass-anisotropy degeneracy introduces a large additional uncertainty
\citep[e.g.][]{Diakogiannis_2014} that limits the utility of the $\beta$ measurements
at present. Instead, the situation is more
promising for GCs in cluster sized halos (which have correspondingly
many more GCs) and \citet{Zhu_2014,Napolitano_2014} and
\citet{Spitler_2012} have inferred anisotropy profiles from these for
M87, NGC~5846 and NGC~1407 respectively, with $M_{200} \approx
10^{13}$~-$10^{15}$~\Msun, which we show in grey curves and symbols
for comparison in Fig.~\ref{fig:beta}. Note that for the Virgo cluster
there is some additional uncertainty in that the cluster may not be
relaxed \citep{Binggeli_1987, Strader_2011}.

Taken at face value, Fig.~\ref{fig:beta} suggests that GCs formed
cosmologically in their own dark matter halos prior to infall into
their (larger) present day hosts would have a larger radial velocity anisotropy than
is observed for real systems. However we caution that one cannot make
strong conclusions due to the size of the uncertainty in observations
and the fact that our predictions are for MW-mass systems and
observational constraints are only available for larger group and
cluster halos. 

Regardless, it seems interesting that at large radii
several observations suggest tangential velocity anisotropy
(i.e. $\beta<0$), which seems difficult to accommodate in this
scenario. In fact, radial anisotropies (especially at large radii)
are not limited to this model only. For instance one would also expect
this if GCs were formed either in the main galaxy and ejected (since
GCs with a distant orbital apocentre would still be expected to have a
near pericentre and thus a nearly radial orbit) and similarly if outer
GCs were stripped from satellites (which would also need to be on
quite radial orbits to be significantly affected by tidal disruption
during close pericentre passages). Mechanisms such as preferential
destruction of GCs on radial orbits can be invoked, but as we saw
earlier this is expected to affect only a modest fraction, and as such
observations suggesting tangential anisotropies beyond 20\% of
$R_{200}$ remain a puzzle. This unusual dynamical arrangement 
of GCs seems reminiscent of the `planes of (accreted) satellites' of M31 \citep{Koch_2006}
- an arrangement that in \lcdm\ is unlikely although not 
exceptional \citep{Cautun_2015}.

\section{The abundance of intergalactic globular clusters}\label{sec:IGCs}

\begin{figure}
\includegraphics[width=\columnwidth]{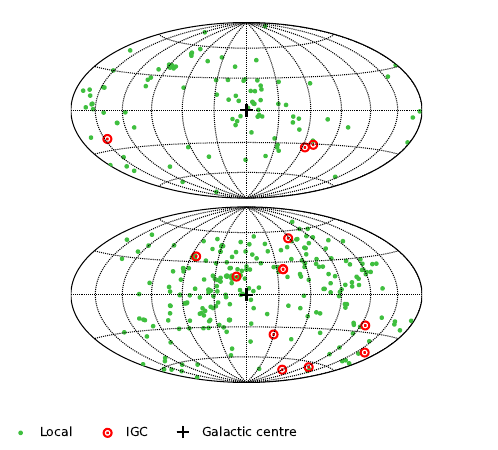}
\caption{Sky distribution of simulated ``intergalactic'' GCs 
  (red dotted circles), defined as those laying within $[1$-$3]
  \;R_{200}$ of a MW-like halo and not bound to any other system with
  $M_{200} > 3 \times 10^8$ ~\Msun\ (i.e. not belonging to any other
  halo of a field dwarf).  We plot the GC distribution from an
  Earth-like vantage: 8~kpc from the centres of the two most massive
  MW-like halos (halos as for Fig.~\ref{fig:orbits}).
  \emph{Green dots} denote (local) ``normal'' candidates within
  $R_{200}$ and \emph{red dotted circles} denote the isolated GC
  candidates. \emph{Black plus} indicates the direction to the halo
  centre. This scenario predicts very few intergalactic GCs, with
  bounds consistent with the current lack of observational candidates
  (see text for more detail).}
\label{fig:IGCs}
\end{figure}

One attractive avenue for the discrimination of GC formation models is
the presence or absence of isolated/free GCs at the present day
floating outside more massive hosts. Since the model where one
populate the high-sigma peaks at high redshift causes GC
progenitors to lie {\it outside} galaxies at their formation, some
fraction of these could still be isolated from galaxies today
(so-called Intergalactic Globular Clusters or IGCs) inhabiting their
own dark matter halo. This is in contrast with models where GCs are
the result of baryonic formation processes in galaxies and essentially
remain associated to the galactic halos for their entire lifetimes
(excluding some almost vanishingly small fraction that are ejected
during halo interactions and whose resulting orbit fails to be
incorporated in either halo). It is important to keep in mind,
however, that the high-sigma peaks at high $z$ are still strongly
clustered, and the vast majority will be incorporated into larger
halos at redshift zero as a result of this (see for instance
Fig.~\ref{fig:projection})

On the observational side the result is a null one, i.e. there are no
observed isolated (i.e. not bound to the halo of a galaxy)
GCs. Marginal cases exist, for example Crater/Laevens I
\citep{Belokurov_2014,Laevens_2014} at $\approx 150 \; \rm kpc$,
however this still puts it inside $R_{200}$ for the MW ($\approx
240$~kpc, e.g. \citealp{Springel08}). Nevertheless the observational
volume over which we can be confident of completeness is
limited. \citet{diTullio_2015} searched the Sloan Digital Sky Survey
(SDSS) and none of the GCs they found appear to be truly isolated
\citep{Mackey_2016}, which from their figures is complete to a depth
of around 750~kpc (around the distance of M31, and smaller than the
$\approx1$~Mpc extent of the Local Group) and the SDSS footprint is
around $14,555^\circ$ (around 30\% of the sky) which gives a volume of
around $0.5\;\rm Mpc^3$.

To evaluate the level of this constraint quantitatively means we would
like to know the abundances of these IGCs inferred from our N-body
simulation. Notably in \lcdm\ the regions from $1$-$3R_{200}$ will on
average be several tens of times denser than the mean density of the
Universe (even excluding that the Local Group is a loose `group'
containing M31, suggesting a density higher than average for a MW mass
galaxy, discussed further in \citealp{Creasey_2015b}) and
correspondingly one could expect the density of GCs to be much higher
too. In order to evaluate this we took the 7 MW-mass halos in our
simulated box and counted GCs that are both in the range
$1$-$3R_{200}$ and \emph{not} incorporated into any other halo above
$3\times 10^8$~\Msun\ ($3\times$ the mass threshold, and thus
significant halo growth usually associated with the merger into a LG
galaxy). A sky-projection for such intergalactic GCs is shown for our
two most massive halos in Fig.~\ref{fig:IGCs} (see red dotted
symbols). These maps assume solar-like position 8~kpc from the centre of the
halo and we have marked for completeness ``normal'' GCs that are
within $R_{200}$ (green dots). Isolated GCs show, as expected, little
clustering.

We find that isolated GCs are indeed very rare, with the most massive
halo having only 9 in this range (lower panel) and the second most
massive only 3 (upper panel). Combining all our data together gives an
abundance measure for GCs between 1 and $3 R_{200}$, denoted $N_{13}$,
of
\begin{equation}
N_{13} = 4.7 \times \left( \frac{M_{200}}{10^{12} \; \rm M_\odot} \right),
\end{equation}
where we have assumed a linear scaling with halo mass and had to apply
a volume correction of a couple of percent due to a slight overlap of
the $R=3R_{200}$ spheres of the seven halos. For comparison we have 
repeated this calculation for the higher redshift tagging ($z=14.1$, $M_{\rm tag}=1.5\times 10^7$~\Msun)
and found this reduces the prefactor by around $40\%$ to $2.77$, i.e. as
one might expect from their higher clustering at high redshift 
that they are less common in the field today. 
The value of $4.7$ for a
fiducial $10^{12}$~\Msun\ MW halo would correspondingly suggest that
for 30\% of the sky (i.e. the SDSS result from \citealp{diTullio_2015}
and \citealp{Mackey_2016}) the expected value would be $1.4$ IGCs and,
assuming Poisson statistics, that the probability of observing zero is
$\exp(-1.4)\approx 25\%$, and easy to accommodate 
(by extension also true for the higher redshift cut). In the future, as
more GCs are identified in the outskirts of our galaxy, the
observation of isolated GCs free floating between galaxies might
become an exciting constraint to the models.

\section{Summary and Conclusions}\label{sec:discussion}

Old globular clusters have ages comparable to the Hubble time, the
remaining uncertainty around which has significant implications for
their formation. A late formation hypothesis ($z \approx 4$) would
indicate a galactic origin (either in the MW or its progenitors and
satellites) whilst an early formation ($z \approx 10$) would be
compatible with they forming at the centres of their own dark matter halos
in the most massive halos at that time (a.k.a high density sigma
peaks, \citealt{Diemand_2005}).

In this work we study various implications of the latter hypothesis
and examine their constraining power when compared to local available
observations. We perform an N-body cosmological simulation of a $10$
Mpc box where we are able to resolve halos with
$M_{200}>10^7$~\Msun. We employ a tagging technique to identify the
GC-forming halos at high redshift and trace their positions to the
present day. In agreement with previous studies, we find that choosing
rare sigma peaks (massive halos) at high redshift as formation sites
for metal poor GCs lead to a highly clustered distribution of GCs
today around more massive galaxies, a simple consequence of the
hierarchical nature of \lcdm.

Interestingly, we find that such simple assumption and tagging process
automatically generates a GC abundance per halo that is almost linear
in halo mass, in good agreement with the observed relation
between the number or mass of GCs and host halo mass
\citep{Hudson_2014}.  The normalisation of that scaling depends
sensitively on the mass of the GC bearing halos and the redshift of
tagging. Once both are fixed, this has consequences for both, the
abundance and the radial distribution of GCs at present day. We find
that using both properties combined can place strong constraints on
this formation scenario for GCs. Namely, early redshifts and higher
masses assumed for tagging result on more centrally concentrated
radial distributions and correspondingly also a lower number of GCs
per host.

If one chooses a mass and redshift cut consistent with the observed
abundance of GCs (in our model this occurs for $M_{\rm
  200}=10^8$~\Msun\ and $z_{\rm tag}=8.65$, adopted as our fiducial model),
the obtained radial distribution is too extended around MW-like hosts
compared to observations. For example, the observed radius containing
half of the MW blue GCs population is $7.1$ kpc, compared to $\sim 54$
kpc median in our fiducial model. Fitting the observed half number
radius of GCs would require us to assume a more strongly biased
population of halos forming GCs at high redshift, which results on a
lower $z=0$ abundance that can account for only 50-60\% the observed
number of metal poor GCs in MW-like systems.  

We conclude that
although this formation scenario might not explain \emph{all} the
observed blue GCs in MW-like halos, it can account for about 50-60\%
of it according to the current observational constraints.  Notice that
some of this tension would be alleviated if the MW and other nearby
galaxies were somehow special, which might be the case if reionization
was inhomogeneous \citep{Spitler_2012}. However, as the radial
distribution of GCs around other MW-mass galaxies have become
available, it appears more unlikely that the MW is an outlier and
consequently may need a different explanation for their rather compact
distribution of metal poor GCs. Encouragingly, this formation scenario
seems to predict the correct abundance and radial extension of metal
poor GCs in dwarf galaxies. 

As such, it is worth exploring further the other predictions of this
formation model in terms of kinematics and distribution of GCs.
Peak-models naturally set up a radial velocity anisotropy where GCs
navigate into the galactic halo on preferentially relatively radial
orbits ($\beta > 0.3$). Counter-intuitively there are a number of
observations that suggest either isotropic or even tangential velocity
anisotropies from tracer material in the halos, albeit the best
evidence comes from larger halos (clusters) and require strong
assumptions of equilibrium and mass distribution. Our N-body
simulations suggest that the intrinsic scatter in the outer halo
(which is primarily caused by substructure) can be quite large,
meaning that the constraining power based on a handful of either
observed or simulated objects is rather weak. Cluster scale larger
volume simulations would be required on the theoretical side.

The high ellipticity expected for GC orbits also means that a large
fraction of GCs must have passed within only a few dozen
kpc of the centre of the host, even for those GCs currently populating the
outskirts of halos. Analytical estimates based on the orbits in our
simulations indicate that the majority of GCs have lost at least half of their initially
bound dark matter mass, and even GCs today at distances as far as
$100$ kpc can be heavily stripped, retaining only 30\% of dark matter
on average but down to 7\% for those in the most highly radial
orbits. This should be kept in mind when attempting dynamical modelling
of the stars in GCs \citep[e.g. ][]{Conroy_2011}. We will
address the consequences for the dark matter content of GCs formed in
this scenario in more detail in a forthcoming paper (Creasey et al.,
{\it in prep}).

One strong constrain to this model is the fact that despite
significant observational efforts, no detection of truly isolated GCs
(i.e. those not associated with the halo of any galaxy) has been made
so far (e.g. \citealp{diTullio_2015,Mackey_2016}). Our model predicts
that intergalactic GCs should be, indeed, extremely rare and considerations of
the volume explored in observations retrieves a predicted detection of
$\approx 1.4$ GC, consistent within Poisson noise with the current null
detection in observational campaigns.

Our results support a scenario where more than half of the blue GCs
could have formed within their own dark matter halos at high redshift
before reionization. Due to the hierarchical and self-similar nature
of substructure within $\Lambda$CDM this simple model provides strong
predictions for the abundance, radial distribution and kinematics of
such population.  

Although the detection of a large dark matter component surrounding
some of these GCs is the most tempting evidence for this formation
scenario, the devil could instead be in the details; and a more
systematic joint study of number, radial distribution and
kinematics of observed GCs around other large systems might prove
equally fruitful at constraining GC formation scenarios. In the MW
itself, the detection of a GC candidate associated to the ultra-faint
dwarf Eri~II prompts the idea of at least some GCs being hosted by
undetected surrounding dark matter halos \citep{Zaritsky2016}. As the
cosmological place of GCs continues to be studied, it must not be
discarded that the oldest and more metal poor GCs could have
originated in the hearts of early collapsed dark matter halos,
constituting the easiest to observe and therefore most accessible
reionization fossils available to our telescopes today.

\section*{Acknowledgements}
The authors would like to thank Marcelo Alvarez, Oleg Gnedin, Raul
Angulo and Alessia Longobardi for useful discussions, Volker Springel
for making {\sc Arepo} available for this work. We would also 
like to thank the anonymous referee for comments which significantly
improved this work. We are grateful to the
hospitality of the Kavli Institute for Theoretical Physics (KITP) at
which some of this work was written, under a research programme
supported in part by the National Science Foundation under Grant
No. NSF PHY11-25915. LVS acknowledges support from the Hellman
Foundation and HST grant HST-AR-14583. OS acknowledges support by NASA 
MUREP Institutional Research Opportunity (MIRO) grant number NNX15AP99A.

\bibliographystyle{mnras}

\bibliography{paper}

\appendix
\section{Tests on GC radial distributions}\label{sec:tests}

\begin{figure}
\includegraphics[width=\columnwidth]{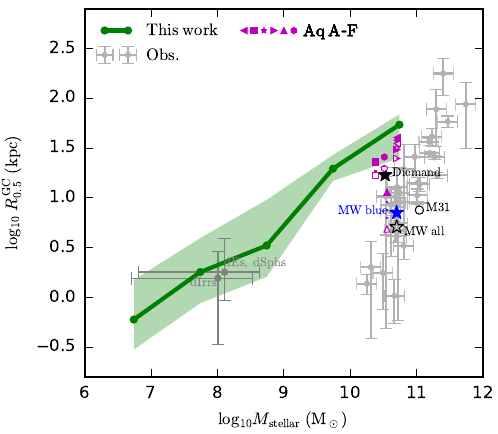}
\caption{ As for Fig.~\ref{fig:halfrad} but exploring the effects of
  adiabatic contraction. \emph{Filled magenta symbols} indicate the Aquarius
  simulations without baryonic contraction, whilst \emph{open symbols} are the 
  median radii including the effect of a $10^{11}$~\Msun\ baryonic disk. 
  \emph{Filled black star} indicates the radius from \citet{Diemand_2005}. 
  Observations (\emph{grey points}) and model (\emph{green}) as for Fig.~\ref{fig:halfrad}.}

\label{fig:halfrad_app}
\end{figure}
One should consider the baryonic effects that we have ignored in this
scenario. Specifically, we have ignored tidal stripping, which will
preferentially remove GCs nearest the centre and \emph{increase} our
median radii (although this will primarily affect the metal-rich
population) and increase the tension. Thus from this perspective, the
radius reported here should be considered \emph{lower} limits to the
expected half-number radii.  On the other hand the baryonic collapse
of material to the centres of halos will contract the orbits and tend
to alleviate the tension. Using the adiabatic approximation for
circular orbits (e.g. \citealp{Blumenthal_1986}) we also indicate the
the contracted radii where we assume a $10^{12}$~\Msun\ halo with a
median concentration of $8.6$. This gives a contraction of
approximately 6~kpc for most of the halos of interest, though this is
probably an overestimate due to the circular orbit approximation
\citep[e.g. ][]{Dutton_2007}. For an easy comparison, we have
indicated the effect of baryonic contraction in our set of Aquarius
halos shown in Fig.~\ref{fig:halfrad}, where filled/open magenta
symbols indicate the results without and with adiabatic contraction,
respectively.

Noticeably in Fig.~\ref{fig:halfrad} the median GC radius from
\citet{Diemand_2005}, which has been averaged over several galaxies,
is more than 1 sigma more compact than our results.  We believe that
this arises due to the different tagging techniques: whereas we tag
only the most-bound particle of a GC-hosting dark matter halo to
trace their $z=0$ position, Diemand et al. uses the distribution of
\emph{all} particles in candidate halos. The latter, as is
equivalent to a mass-weighted average that is mostly tracing the main
progenitor of a MW halo, instead of the many more accreted and smaller
subhalos. As an estimate, a $10^{12}$~\Msun\ halo at $z=0$ can have a
main progenitor with mass $10^{10}$ already at $z=9$, dominant over the
contribution from the other subhalos.
Other causes for the difference include the cosmology and the unique
formation history of each
galaxy.

Another work that has found more compatible (compact) predicted GC
radii is \cite{Griffen_2010}, for a set of GCs for the Milky-Way
analogue Aquarius A-2. This required the use of a relatively high
reionization redshift for the MW ($z\approx 13$), and so the GCs are
more clustered. We tested our method on the Aquarius
\citep{Springel08} halos (these are of course a slightly different
cosmology) and as see in Fig.~\ref{fig:halfrad_app} they fall, 
within the range of our distribution, with the exception of
Aquarius~E-4\footnote{Usually one expects Aq-F to be the outlier since
  it experiences a late major merger, but surprisingly this not the
  case here.}. E-4 is the halo
with the \emph{least} tagged halos, which seems to be correlated
with more compact distributions in our simulations.

A final test we made was to consider GC formation not only at the 
tagging redshift but prior redshifts as well, comparable for example to
\citet{Griffen_2010}. As such we ran an additional high time-resolution 
simulation down to redshift $8.65$ in order to identify how many of our
halos of mass above  $M_{\rm tag}$ had in fact achieved this mass in more than
one progenitor, however this only appeared to increase the number of halos by around
3\%, and as such would make little difference to our result.

\end{document}